\documentclass[preprint,onecolumn,amsmath,amssymb,floatfix]{revtex4}
\usepackage{graphicx}

\usepackage{times}

\begin{document}

% The following parameters seem to provide a reasonable page setup.

%\topmargin 0.0cm
%\oddsidemargin 0.2cm
%\textwidth 16cm 
%\textheight 21cm
%\footskip 1.0cm

\title{Simultaneous measurements of electronic conduction and Raman response in molecular junctions}

\author{Daniel R. Ward$^{1}$, Naomi J. Halas$^{2,3,5}$, Jacob W. Ciszek$^{6}$, James M. Tour$^{3}$, Yanpeng Wu$^{4}$, Peter Nordlander$^{1,5}$, Douglas Natelson$^{1,2,5}$}

\affiliation{$^{1}$Department of Physics and Astronomy, 
$^{2}$Department of Electrical and Computer Engineering, 
$^{3}$Department of Chemistry,
$^{4}$Applied Physics Graduate Program,
$^{5}$Rice Quantum Institute, Rice University, 6100 Main St., Houston, TX  77005, USA
$^{6}$Department of Chemistry, Northwestern University, Evanston, IL 60208, USA
}

\date{\today}

\begin{abstract}
Electronic conduction through single molecules is affected by the
molecular electronic structure as well as by other information that is
extremely difficult to assess, such as bonding geometry and chemical
environment.  The lack of an independent diagnostic technique has long
hampered single-molecule conductance studies.  We report simultaneous
measurement of the conductance and the Raman spectra of nanoscale
junctions used for single-molecule electronic experiments.  Blinking
and spectral diffusion in the Raman response of both
\textit{para}-mercaptoaniline and a fluorinated oligophenylyne
ethynylene correlate in time with changes in the electronic
conductance.  Finite difference time domain calculations confirm that
these correlations do not result from the conductance modifying the
Raman enhancement.  Therefore, these observations strongly imply that
multimodal sensing of individual molecules is possible in these
mass-producible nanostructures.
\end{abstract}

\maketitle

The molecular-scale limits of electronic conduction are of fundamental
scientific interest and relevant to future technologies.  Our
understanding of electronic conduction through {\em single} small molecules
has grown dramatically in the last decade thanks to improved
techniques, including mechanical break
junctions\cite{ReedetAl97Science,HeetAl06FD,VenkataramanetAl06NL}
single-molecule transistors
(SMTs)\cite{ParketAl00Nature,ParketAl02Nature,LiangetAl02Nature,KubatkinetAl03Nature,YuetAl04NL,PasupathyetAl04Science,YuetAl04PRL,PasupathyetAl05NL,ChampagneetAl05NL,YuetAl05PRL,HeerscheetAl06PRL,ChaeetAl06NL,DanilovetAl06NL,vanderZantetAl06PSSB,NatelsonetAl06CP,DanilovetAl07NL},
nanoparticle dimers\cite{DadoshetAl05Nature}, noise
characterization\cite{DjukicetAl06NL}, and thermopower
measurement\cite{ReddyetAl07Science}.  A major complication in
interpreting these experiments is the lack of local imaging or
spectroscopic tools that can assess the environment and presence of
the molecule of interest.

Over the same period, single-molecule spectroscopies have progressed
substantially.  In particular, surface-enhanced Raman spectroscopy
(SERS) has been demonstrated with single-molecule
sensitivity\cite{KneippetAl97PRL,NieetAl97Science,XuetAl99PRL,MichaelsetAl00JPCB}
in random aggregates of metal nanoparticles, though this level of
detection is very challenging to prove conclusively.  The
electromagnetic component of SERS enhancement results from the
excitation of surface plasmons in the metal, leading to local fields
at the molecule enhanced by a factor of $g(\omega)$ relative to the
incident field.  The Raman cross section is then enhanced by
$g(\omega)^{2}g(\omega')^{2}$, where $\omega$ and $\omega'$ are the
frequencies of the incident and Raman-scattered radiation,
respectively.  Electromagnetic SERS enhancements exceeding $10^{12}$
are needed to approach single-molecule detection for many
molecules without resonant Raman effects\cite{LeRuetAl07JPCC}.  Additional
``chemical'' enhancement is also possible due to electronic
interactions between the molecule and the metal.  The vibrational
modes observed in SERS reflect both the molecule and the local
environment and conformation of the molecule on the metal surface.
Using a sharp tip to provide a very large local field enhancement,
Raman sensitivity has approached the single-molecule
level\cite{NeascuetAl06PRB,DomkeetAl06JACS}.  Very recently we found
that nanoscale gaps between extended electrodes are very effective as
extremely confined SERS hotspots and may be mass produced with
relatively high yields.\cite{WardetAl07NL}

In this paper we use these nanoscale gap structures to perform
simultaneous measurements of electronic transport and SERS.  In many
previous
papers\cite{ParketAl00Nature,ParketAl02Nature,LiangetAl02Nature,KubatkinetAl03Nature,YuetAl04NL,PasupathyetAl04Science,YuetAl04PRL,PasupathyetAl05NL,ChampagneetAl05NL,YuetAl05PRL,HeerscheetAl06PRL,ChaeetAl06NL,DanilovetAl06NL,vanderZantetAl06PSSB,NatelsonetAl06CP,DanilovetAl07NL}
it has been established that conductance in such structures is
dominated by roughly a molecular volume.  The conductance as a
function of time is observed to correlate strongly with the SERS
signal in 11\% of the junctions measured.  Conductance changes
correlate with sudden changes in the intensity of sets of Raman modes
(``blinking'') and with spectral diffusion of mode positions.  The
data suggest that both SERS and conductance changes are most likely
due to changes in conformation and binding of an individual molecule.
The combined data provide a great deal of information about the effect
of molecular orientation and environment on both conduction and SERS,
although a detailed understanding of this correlated information is
indeed a very significant theoretical challenge.  The most likely
explanation for these results is that single-molecule multimodal
sensing is possible.  This combined measurement technique also opens
the possibility of direct assessment of vibrational pumping and local
heating in single-molecule electronic transport.

Our nanogap structures are fabricated on $\sim$ 1~cm$^{2}$ pieces of
$n+$-doped Si wafer with 200 nm of thermal silicon oxide.  The
structures are defined using electron-beam lithography and e-beam
evaporation of 1~nm Ti and 15~nm of Au.  The initial nanoconstriction
structure consists of two large pads connected by a single
constriction as shown in Figure 1.  The constriction is approximately
500 nm long and 100-180 nm wide.  Liftoff is performed in acetone or
chloroform and the samples are then cleaned of organic residue by 1
minute exposure to O$_2$ plasma.  For devices incorporating
para-mercaptoaniline (\textit{p}MA), samples are immediately placed in
a 1~mM solution of \textit{p}MA in ethanol. Samples are soaked in the
\textit{p}MA solution for 12 to 24 hours as \textit{p}MA
self-assembles on the Au surfaces, followed by an ethanol rinse to
remove any excess \textit{p}MA.  We have also measured devices using a
fluorinated oligophenylene ethynylene (FOPE)\cite{HamadanietAl06NL}
that possesses a distinctive Raman spectrum (see Supporting
Information).  FOPE was assembled via standard base deprotection\cite{TouretAl95JACS} from a 0.25~mg/mL solution of the
thioacetate form of the molecules in 1:1 ethanol-chloroform solvent,
prepared under dry N$_{2}$ gas.

The nanoconstrictions are converted into nanogaps via
electromigration\cite{ParketAl99APL}, a thoroughly
studied\cite{StrachanetAl06NL,TaychatanapatetAl07NL} process dominated
by momentum transfer from current-carrying electrons to mobile atoms
in the metallic lattice.  Electromigration has been used extensively
to prepare electrodes for single-molecule conduction measurements,
with typical yields of $\sim$~10-20\% for tunneling gaps inferred to
contain individual molecules (based on statistics on thousands of
junctions)\cite{ParketAl00Nature,ParketAl02Nature,LiangetAl02Nature,YuetAl04NL,PasupathyetAl04Science,YuetAl04PRL,PasupathyetAl05NL,ChampagneetAl05NL,YuetAl05PRL,HeerscheetAl06PRL,ChaeetAl06NL,vanderZantetAl06PSSB,NatelsonetAl06CP}.
Each gap is electromigrated with an automated procedure to form an
atomic-scale constriction with a resistance of $\sim$3~k$\Omega$,
which is then allowed to break spontaneously\cite{O'NeilletAl07APL}.
While the atomic-scale details of each gap are different, gaps with
measurable tunneling currents are formed routinely with high yield,
and recent advances in {\em it situ} electron
microscopy\cite{StrachanetAl06NL,HeerscheetAl07APL} do, in principle,
permit detailed structural examinations of the resulting electrodes.
The migration and subsequent electrical measurements are performed
\textit{in situ} on the sample stage of the Raman measurement system,
in air at room temperature.

Electrical contact to the junction under test is made via
micropositionable probes.  One digital lock-in amplifier (SRS SR830)
is used to source 50-100~mV RMS at 200~Hz onto one pad, while the other
pad is connected to a current-to-voltage converter (either SRS SR570
or Keithley 482).  The AC current ($\propto dI/dV$) and its second
harmonic ($\propto d^{2}I/dV^{2}$) are measured with lock-in
amplifiers, while the DC component of the current is sampled at
5.0~kHz.  The unusually large AC bias (much larger than necessary to
measure differential conductance alone) is required because of an
unanticipated complication: the illuminated, molecule-decorated
nanogaps can also exhibit significant DC photocurrents due to optical
rectification (to be described in a separate publication).  The large
AC bias is needed so that the AC current is detectable without the DC
current signal overloading the lock-in input stages.  We find no
evidence that the 100~mV RMS AC bias degrades the nanogap or the
assembled molecules.

Optical measurements are performed using a WITec CRM 200 scanning
confocal Raman microscope in reflection mode.  Devices are illuminated
by a 785 nm diode laser at normal incidence with a diffraction-limited
spot.  A 100x ultra-long working distance objective leaves sufficient
room for the micromanipulated electrical probes to be inserted between
the objective and sample.

As reported previously, the electromigrated nanogaps are outstanding
substrates for SERS.\cite{WardetAl07NL} Initially, spatial maps of the
underlying Si of unmigrated nanogaps are obtained to facilitate
centering of the Raman microscope over the nanogap to within 100 nm.
Spatial maps of the integrated molecular Raman signal after
electromigration demonstrate the localization of the SERS hotspot
(Fig.~1D, E).  Raman spectra are taken with 1~s or 2~s integration
times while the microscope objective is held fixed over the migrated
junctions.  Electrical measurements on unbroken constrictions under
various illumination conditions demonstrate that heating of the
electrodes due to the laser is not significant.  The inferred change
in the electrode temperature at $\sim$ 0.5~mW laser power was less
than 2~K (see Supporting Information).

When the conductance of the junction drops below the conductance
quantum, $G_0\equiv 2e^2/h$, a tunneling gap is formed, and
simultaneous conductance and Raman measurements are performed.
\textit{In situ} measurements of the optical response of nanogaps
during migration are presented in Figure 2.  Even prior to complete
nanogap formation, partially electromigrated junctions show SERS
enhancement of the assembled molecules once the resistance exceeds
about 1~k$\Omega$.  The appearance of SERS indicates that the local
interelectrode plasmon modes are now excitable.  Crudely, this implies
that over an optical cycle the junction acts more like a capacitor
than a resistor; that is, the $RC$ time constant of the nanogap is
comparable to one optical period.  For a 1~k$\Omega$ nanogap
illuminated at 785~nm this implies an effective nanogap capacitance at
optical frequencies on the order of 10s of attofarads.

The measured Raman signal strength scales logarithmically with the
resistance of the gap until resistances exceed 1-10~M$\Omega$.  At
higher gap resistances, the Raman signal takes on a roughly constant
value with sporadic changes (corresponding to SERS blinking events).
This decoupling of electronic transport and SERS at low conductances
is not surprising \textit{a priori}, since tunneling conductances vary
exponentially with gap size, while local plasmonic structure is less
sensitive.  Blinking events are often not correlated with further
changes in junction resistance.  This means that molecules are present
in a region of strong Raman enhancement, while the molecular-scale
tunneling volume dominating interelectrode conductance does not
contain molecules that are detectably contributing to the Raman
signal.  We discuss this further below.

However, in 17 of 120 junctions using \textit{p}MA and 4 out of 70 junctions using FOPE, there are strong temporal correlations
between the fluctuations in the nanogap conductance and changes in the
SERS spectrum.  This yield is \textit{quantitatively consistent} with
the yield of tunneling gaps containing single molecules inferred in
single-molecule transistor measurements.  Examples are shown in
Figures 3 and 4 (with further examples in the Supporting Information).
In Figure 3A a simple positive correlation between Raman intensity and
$dI/dV$ is observed for all Raman modes.  In Figure 3B another
positive correlation between Raman intensity and differential
conductance is observed in a different junction.  In this case
spectral diffusion of the Raman lines occurs but does not correlate
significantly with the conduction.  In both Figs. 3A and 3B the
amplitudes (count rates) of strong Raman modes have similar relative
changes as the differential conductance.

Figure 4 is an example of a more complicated relationship between the
conductance and the SERS spectrum.  While sudden changes in the Raman
spectrum are correlated in time with changes in the measured
conductance, some increases in Raman intensity correlate with
increased conductance, while others correlate with decreases in
conductance.  Additionally, changes in the mode \textit{structure} of
the Raman spectrum clearly correlate with changes in
conductance.  In region A the Raman spectrum and conductance are
constant; when the Raman spectrum changes in region B a three-fold
increase in conductance is observed, though overall Raman intensity
changes only for certain modes.  In region C the spectrum changes yet
again, and while a weaker Raman spectrum remains, the conduction drops
significantly.  In region D the \textit{same} mode structure seen in B
\textit{returns} and the conduction is similar to that seen in B as
well.  Regions D and E have positive correlations between Raman
intensity and conduction with the lowest conduction observed between D
and E where the Raman intensity is also lowest.  At F a switch from
positive to negative correlation between the Raman intensity and
conduction occurs and carries over to regions G and H.  At region I a
small change in the mode structure is observed correlating with a
switch to positive intensity-conduction correlations, continuing
through regions J and K.  In region L there is a final change in mode
structure resulting in negative intensity-conduction correlations
exemplified in the three conduction spikes that occur when the Raman
spectrum disappears.

It should be noted that our \textit{p}MA spectra are typically
dominated by the $b_2$ symmetry modes\cite{Osawa94JPhysChem}, as was
seen previously\cite{WardetAl07NL}.  This is not surprising, as it is
well accepted that $b_2$ symmetry modes experience additional
''chemical'' enhancements in comparison to $a_1$ symmetry modes.  We
often only observe the 1590 cm$^{-1}$ $a_1$ symmetry peak and not the
other expected $a_1$ mode at 1077 cm$^{-1}$.  Strong spectral
diffusion in both molecules with shifts as large as $\pm$20 cm$^{-1}$
have also been observed, and are clear in Fig. 3b.  This surely limits
direct comparison to spectra reported elsewhere.  However, the
measured spectra are quite consistent with one another and are
qualitatively different than those seen in ``bare'' junctions
contaminated by physisorbed exogenous carbon\cite{WardetAl07NL}.

One possible concern could be that changes in metal configuration at
the junction are responsible for the fluctuations in tunneling
conductance and SERS intensity.  This scenario is unlikely for several
reasons.  First, tunneling conductances depend exponentially on gap
geometry; while $dI/dV$ could change by a factor of ten for a 0.1~nm
change in gap separation, it is very unlikely that the electromagnetic
enhancement would be as strongly affected.  Second, it is not clear
how metal rearrangement could explain the observed changes in Fig.~4;
this would require that the gap itself alternately grow and shrink,
with some changes in metal geometry giving large $dI/dV$ features with
small Raman effects and others vice versa.  Finally, during the events
in Figs. 3 and 4, the continuum emission at low wavenumbers observed
previously due to inelastic light scattering from the metal
electrodes\cite{WardetAl07NL} is constant in time (see Supporting
Information).

Another concern is that changes in tunneling conduction in one part of
the junction may alter the plasmon mode structure and affect Raman
emission from elsewhere in the junction.  Such a scenario could lead
to correlations like those in Figs. 3 and 4 even if conduction and
Raman emission are not from the same molecule. Given that the
interelectrode conductance affects Raman emission (Fig.~2), it is
important to consider this possibility.  We have performed
finite-difference time domain (FDTD) simulations of the optical
properties of such junctions to assess this issue, and the results
({\it vide infra}) effectively rule out this concern.  While the
finite grid size (1~nm) required for practical computation times
restricts the quantitative accuracy of these calculations, the main
results regarding spatial mode structure and wavelength dependence are
robust, and the calculated electric field enhancements are an {\it
  underestimate}\cite{OubreNordlander05JPCB,WardetAl07NL}.

Figure~5 shows a comparison of calculated extinction spectra that
characterize the plasmonic mode structure of the gap structure shown,
for various values of interelectrode conductance connecting the source
and drain at the indicated point.  An analysis of the instantaneous
charge distribution associated with the plasmon resonances in Fig.~5
shows that negligible charge transfer occurs between the two
electrodes for conductances smaller than $G_{0}$.  The FDTD
simulations show that the mode structure and enhancement are {\it
  unaffected} by conductances smaller than a few $G_{0}$.  Details are
presented in Supporting Information.  These calculations confirm the
interpretation given above for Fig. 2: The plasmonic mode structure
responsible for enhanced local fields in the nanogap is established
once the interelectrode conductance falls well below $G_{0}$.  Given
these FDTD results, the only plausible explanation for such strong
correlations in time between conduction and Raman emission is that
both processes involve the same molecule or molecules.

Conduction in nanogaps is known to be dominated by a tiny volume
inferred in single molecule transport and break junction experiments
to contain often only one molecule.  Molecular movement, changes in
bonding, or reorientation of the molecule in the gap results in
different tunneling configurations and hence in conductance changes.
The complex relationship between conductance and Raman mode structure
and intensity is also then natural, since chemical enhancement effects
and the appearance of broken symmetry $b_{2}$ modes should be strongly
affected by changes in molecular configuration on the metal surface.
It should also be noted that the measured junction conductances are
consistent with the expected single molecular conductance range of $10^{-3}$
to $10^{-4}$ G$_0$ measured in similar molecules by break junction
techniques\cite{VenkataramanetAl06NL}.

The detailed SERS mode structure combined with the conductance
contains a wealth of information about the bonding, orientation,
and local environment of the molecule.  With appropriate
electronic structure calculations and theoretical estimates of
the Raman tensor for candidate molecule/metal configurations,
it should be possible to infer likely junction geometries and
chemical structure corresponding to each type of Raman spectrum.
Such calculations are very challenging even for the conductance
distribution alone\cite{QueketAl07arxiv}.

These conductance/Raman observations and accompanying calculations
demonstrate that electromigrated nanogaps between extended electrodes
can achieve enhancements sufficient for single-molecule SERS
sensitivity.  Given that these structures can be fabricated in a
scaleable manner in predefined locations with high
yields\cite{WardetAl07NL}, this may allow significant advances in
SERS-based sensing as well as multimodal sensing.  With further
improvements in technique (\textit{e.g.}, measurements in vacuum as a
function of temperature, interelectrode bias, and gate voltage in a
SMT configuration), it will be possible to address open fundamental
issues in SERS, including the nature of chemical enhancement, the
mechanism of blinking, and the cause of the large spectral diffusion
of Raman lines.  Finally, comparisons of Stokes and anti-Stokes Raman
peak intensities as a function of bias across the junction can reveal
whether current flow pumps particular vibrational modes out of thermal
equilibrium.  This would enable new and detailed studies of
nonequilibrium physics and chemistry at the single molecule scale.
Comparisons between these results and transport/Raman studies on
molecular ensembles\cite{NowaketAl04JACS,JaiswaletAl06AC} should also
provide valuable information about the effect of molecular environment
on these processes.

DW acknowledges support from the NSF-funded Integrative Graduate
Research and Educational Training (IGERT) program in Nanophotonics.
NH, DN, and PN acknowledge support from Robert A. Welch Foundation
grants C-1220, C-1636, and C-1222, respectively.  DN also acknowledges
NSF award DMR-0347253, the David and Lucille Packard
Foundation, the Sloan Foundation, and the Research Corporation.
JMT acknowledges support from DARPA and AFOSR.

{\bf Supporting Information Available:}  Detailed examination of laser heating, continuum background, extended discussions of FDTD calculations, and more example data sets.  This material is available free of 
charge via the Internet at http://pubs.acs.org.

\clearpage

\begin{figure}[h!]
\begin{center}
\includegraphics[clip, width=7in]{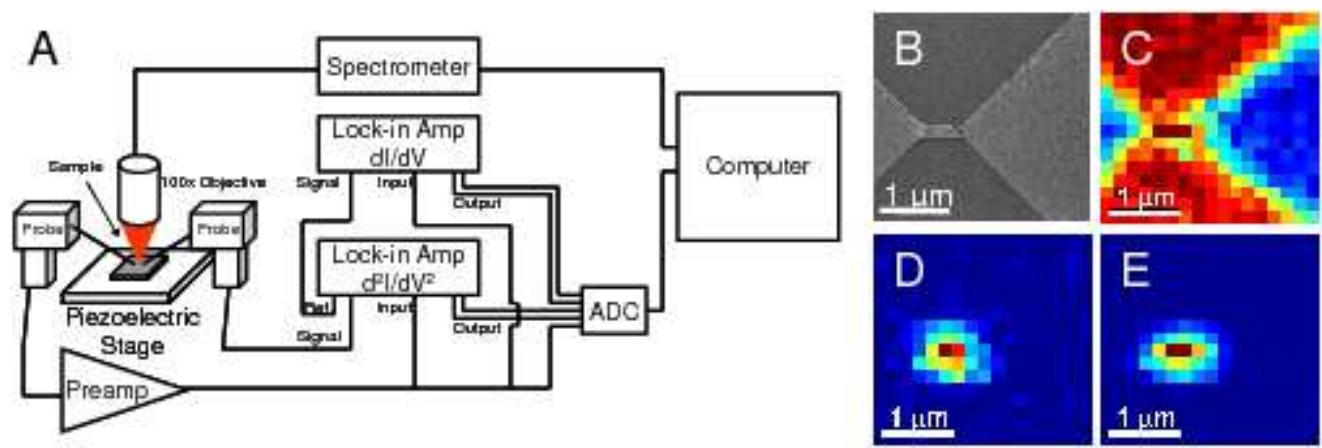}
\end{center}
%\vspace{-5mm}
\caption{\small (A) Schematic of the electronic measurement.  A 100~mV RMS AC signal is sourced by a lock-in into one pad.  The AC current and its second harmonic are measured by lock-in amplifiers.  The DC current is sampled with a current-to-voltage amplifier at 5~kHz. Raman spectra are synchronously captured with 1-2~s integrations at an incident 785~nm wavelength laser intensity of about 0.5~mW.  
(B) Scanning electron image of Au constriction with nanogap.  The constriction is 180 nm wide with a gap $<$~5~nm in size.
(C) Map of the substrate Si 520 cm$^{-1}$ peak (integrated from 480-560 cm$^{-1})$ of the device from B.  Red corresponds to the highest number of CCD counts and blue is the fewest counts.  The Au pads which attenuate the Si signal are clearly visible.
(D) Map of the \textit{p}MA SERS signal from device in B from the $a_{1}$ symmetry mode at 1590 cm$^{-1}$ (integrated from 1550-1650 cm$^{-1}$), showing that the Raman signal is localized only to the nanogap region.
(E) Map of integrated continuum signal (due to inelastic light scattering from the metal electrodes) from device in B (integrated from 50-300 cm$^{-1}$).}
\label{fig1}
%\vspace{-5mm}
\end{figure}

\clearpage

\begin{figure}[h!]
\begin{center}
\includegraphics[clip, width=3.25in]{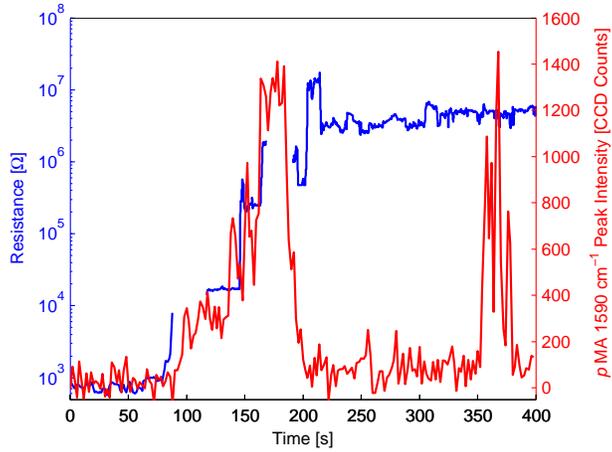}
\end{center}
%\vspace{-5mm}
\caption{\small 
Blue curve (left scale): Resistance as a function of time for a nanogap as it is migrated.  Migration is complete when the resistance reaches $\sim$ 13 k$\Omega \approx (1/G_{0})$.  Breaks in the curve occur where the gain of the current amplifier was being changed to maintain signal.  Red curve (right scale): CCD counts per second in the 1590 cm$^{-1}$ peak (integrated from 1550-1650 cm$^{-1}$) as a function of time for the same device (synchronized with resistance plot).  The intensity of the peak increases linearly with the log of  the resistance until the resistance reaches around $10^{6}~\Omega$, at which point the intensity drops significantly and no longer shows correlations with the resistance.  Stochastic intensity fluctuations (``blinking'') are observed beyond this point. } 
\label{fig2}
%\vspace{-5mm}
\end{figure}

\clearpage

\begin{figure}[h!]
\begin{center}
\includegraphics[clip, width=5in]{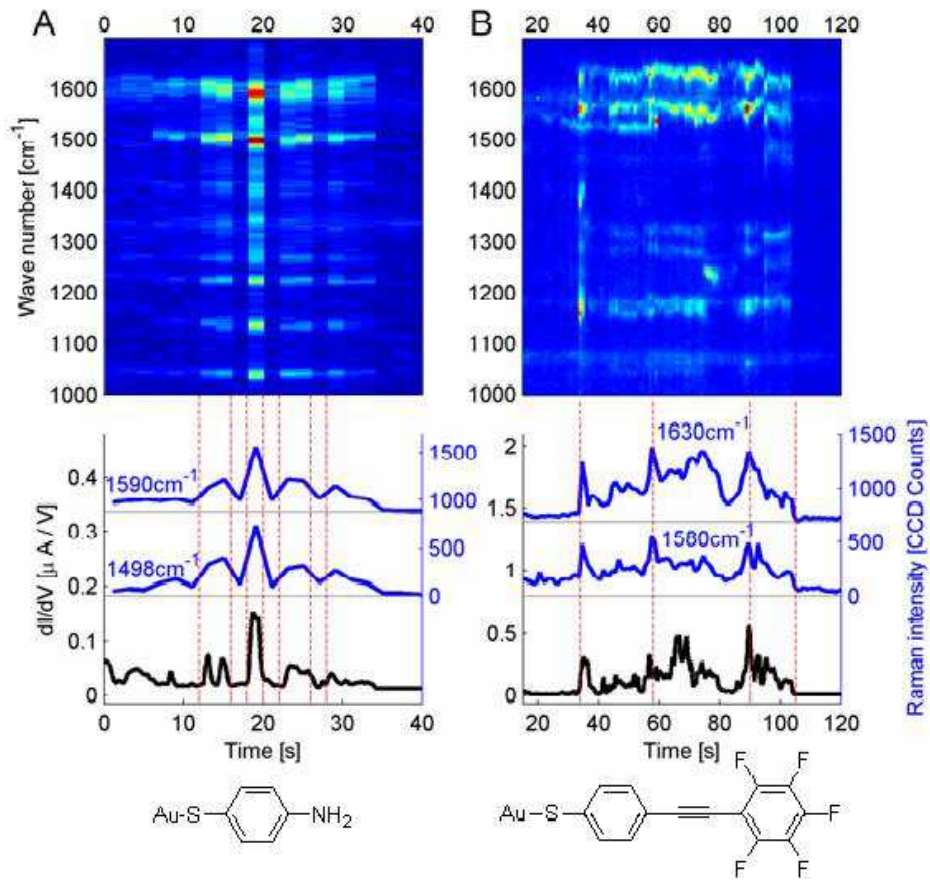}
\end{center}
%\vspace{-5mm}
\caption{\small (A) Waterfall plot of Raman spectrum (2~s integrations) and positively correlated differential conductance measurements (dark blue = 50 counts; dark red = 160 counts) for a \textit{p}MA sample.  All Raman modes that are visible exhibit this behavior as illustrated by the 1498 cm$^{-1}$ and 1590 cm$^{-1}$ modes.  The 1590 cm$^{-1}$ mode has been shifted upward on the lower graph with the gray line indicating zero CCD counts.  Vertical red lines indicate points of rapidly changing Raman intensity and conduction. Structure of \textit{p}MA after self-assembly onto Au is shown below.
(B) Waterfall plot of Raman spectrum (1 s integrations) and positively correlated conductance measurement for a FOPE sample (dark blue = 0 counts; dark red = 250 counts). Strong spectral wondering is observed with no correlation to changes in conductance. Both visible modes at 1580 cm$^{-1}$ and 1630 cm$^{-1}$ are positively correlated.  The slower response of the Raman spectrum compared to the conductance is due to the relatively long integration time. The 1630 cm$^{-1}$ mode has been shifted upward on the lower graph for clarity. Structure of FOPE after self-assembly onto Au is shown below.
}
\label{fig3}
%\vspace{-5mm}
\end{figure}

\clearpage

\begin{figure}[h!]
\begin{center}
\includegraphics[clip, width=5in]{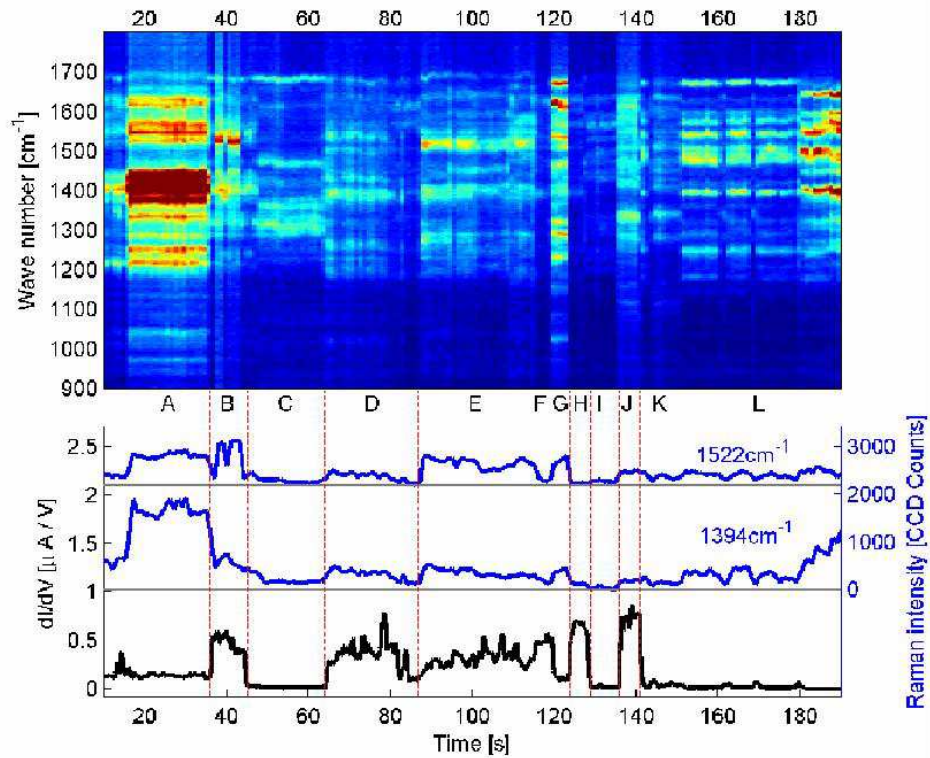}
\end{center}
%\vspace{-5mm}
\caption{\small Waterfall plot of Raman spectrum (1 s integrations) and conduction measurements for a \textit{p}MA sample.  The device experiences periods of correlation (regions B,D,E) and anticorrelation (region L) between Raman intensity and conduction.  Distinct changes in conduction are observed with every significant change in the Raman spectrum and are indicated by vertical red lines.  The modes near 1394 cm$^{-1}$ and 1522 cm$^{-1}$ show similar intensity fluctuations except at region B and the end of region L.  The color scale (dark blue = 20 counts; dark red = 200 counts)has been set to make as many Raman modes visible as possible visible.  This results in the saturation of the signal at region A which would otherwise resolve into well defined peaks. The 1522 cm$^{-1}$ mode has been shifted upward on the lower graph for clarity.
}
\label{fig4}
%\vspace{-5mm}
\end{figure}

\clearpage

\begin{figure}[h!]
\begin{center}
\includegraphics[clip, width=5in]{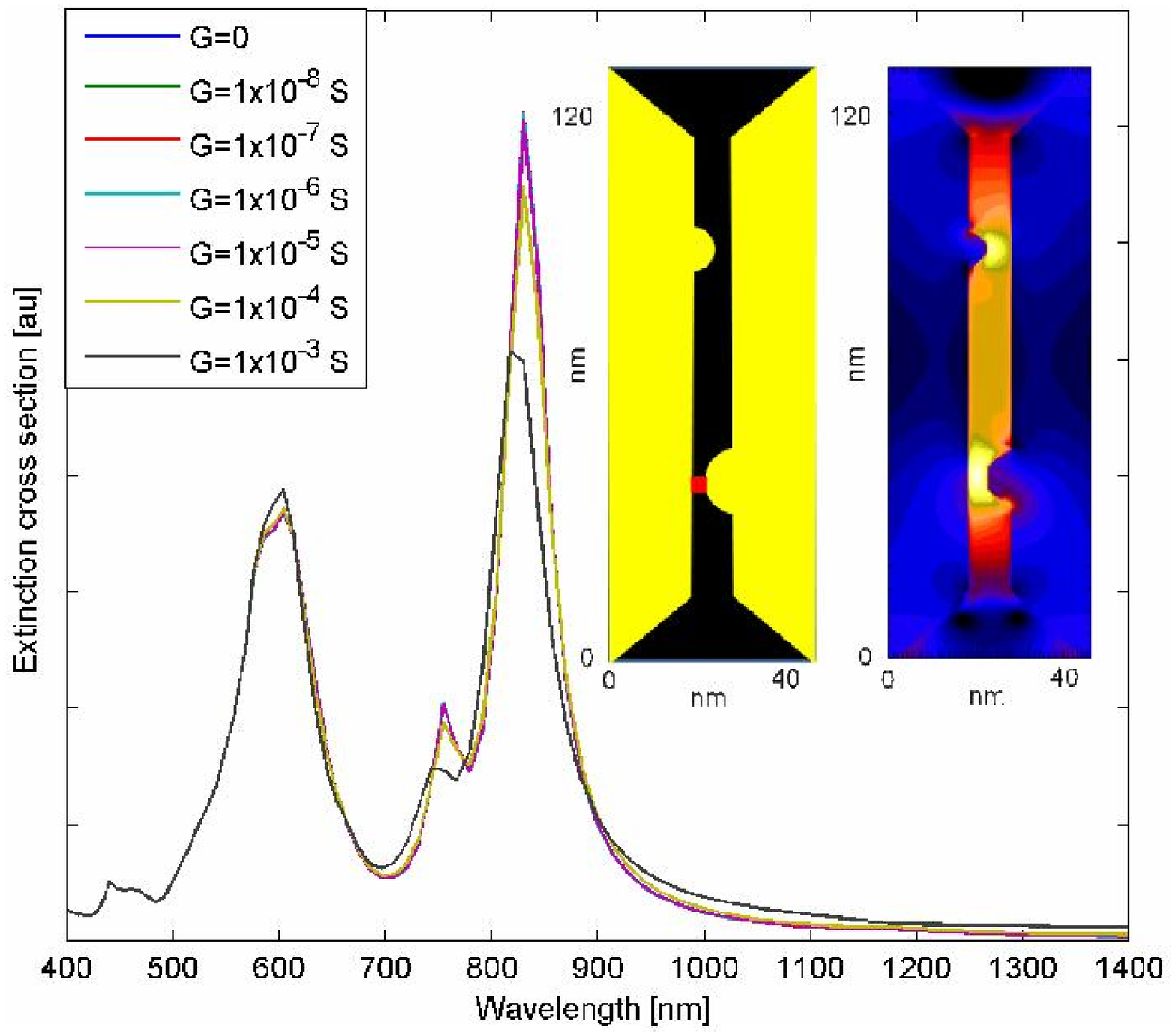}
\end{center}
%\vspace{-5mm}
\caption{\small Extinction spectrum calculated using a 1~nm grid size for the structure partially shown in the left inset.  The electrodes are modeled as Au, 15~nm thick, sitting on 50~nm thick SiO$_{2}$ dielectric, with an overall interelectrode gap of 8~nm.  The upper and lower protrusions into that gap shown are modeled as hemispheres of radius 4~nm and 6~nm, respectively.  The red square indicates the location of the modeled interelectrode conductance (a volume 2~nm on a side, meant to represent a molecule at the interelectrode gap).  The right inset shows a map of $|{\mathbf E}|^{4}$, where ${\mathbf E}$ is the local electric field normalized by the magnitude of the incident field (roughly the Raman enhancement factor), for the mode near 825~nm. White corresponds to an enhancement of $10^{9}$.  This field map is essentially unchanged until the junction conductance approachs $10^{-4}$~S $\sim G_{0}$. }
\label{fig5}
%\vspace{-5mm}
\end{figure}

\clearpage

{\center{\bf Supporting Information: \\ Simultaneous measurements of electronic conduction and Raman response in molecular junctions}}

\renewcommand{\thefigure}{S\arabic{figure}}
\setcounter{figure}{0}

\section{Laser Heating}
To determine the effects of laser heating on nanogaps, unbroken constrictions were illuminated with various laser powers while the resistance was measured as seen in Figure S1A.  To determine the laser power we measured the number of CCD counts per second in the Si 520~cm$^{-1}$ peak.  At higher laser powers irreversible changes to the constriction occur as evident in the dropping resistance of the constriction when the laser power is held constant.  We do not believe that such changes in the constriction will dramatically change our estimate of the laser heating.  These laser powers are much higher than used in our conduction experiment where we expect about 50 CCD counts per second in the Si peak.    From our measurements we determined the resistance of the constriction as a function of laser power as shown in Figure S1B using a linear least-squares fit.  The final resistance/laser intensity relationship of $R=1.35 \times 10^{-3} I_{laser} + 180~\Omega$, where $R$ is the resistance in Ohms and $I_{laser}$ the laser intensity in Si CCD counts/second, is based on the average of three different samples.  

We then took different devices from the same sample chip and heated them in a probe station while measuring their resistance, as seen in Figure S1C.  A linear fit was performed to determine the resistance as a function of temperature this time averaging over four samples.  The final relationship was determined to be $R=0.11T+85~\Omega$, where $R$ is the resistance in Ohms and $T$ is the temperature in Kelvin.  Equating our two relationships we find that the change in temperature per CCD count in the Si peak is 0.012~K.  At the laser power used in our conduction experiments (50 CCD counts/second) this lets us estimate a temperature change of 0.62~K due to illumination.  

\begin{figure}[ht]
\begin{center}
\includegraphics[clip, width=5in]{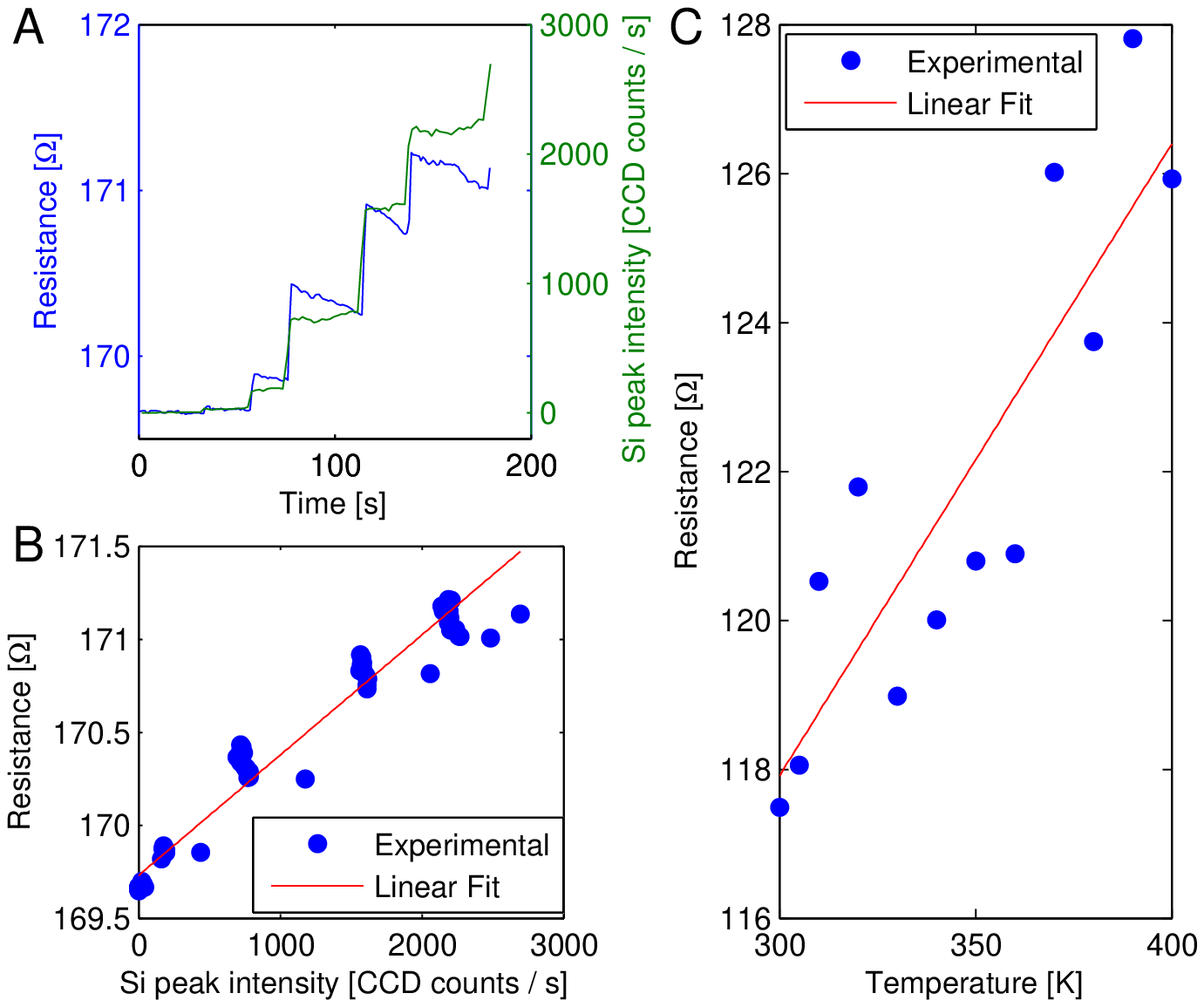}
\end{center}
\caption{
(A) Resistance and laser intensity (in Si peak CCD counts/second) for an illuminated constriction as a function of time.  It is clear that the resistance scales linearly with laser power.  (B) Resistance as a function of laser intensity (again in Si peak CCD counts/second) for the same device as in A.  The data is well represented by a linear fit of $R=0.64 \times 10^{-3}I_{laser}+170~\Omega$ where $R$ is the resistance in Ohms and $I_{laser}$ is the laser intensity in CCD counts/second.  (C) Resistance as a function of temperature for a different constriction heated in a vacuum probe station.  The data is fit to the line $R=0.085T+92~\Omega$ where $R$ is the resistance in Ohms and $T$ is the temperature in Kelvin.
}
\end{figure}

\clearpage

\section{Additional examples}
Figure S2 is a second example of the evolution of SERS response 
during the electromigration process.  As in Fig.~1 of the main 
manuscript, Raman response becomes detectable at the nanojunction 
once the electromigrated two-terminal resistance exceeds 1-2~k$\Omega$.
This implies that the localized interelectrode plasmon modes 
responsible for the enhanced local electric field are established
before the junction is entirely broken.  As discussed in the main
text, a simple interpretation of this is that once the charge 
relaxation time across the nanoconstriction significantly 
exceeds the duration of an optical cycle, the coupling between
the two sides of the junction is chiefly capacitive rather than
conductive.

\begin{figure}[ht]
\begin{center}
\includegraphics[clip, width=5in]{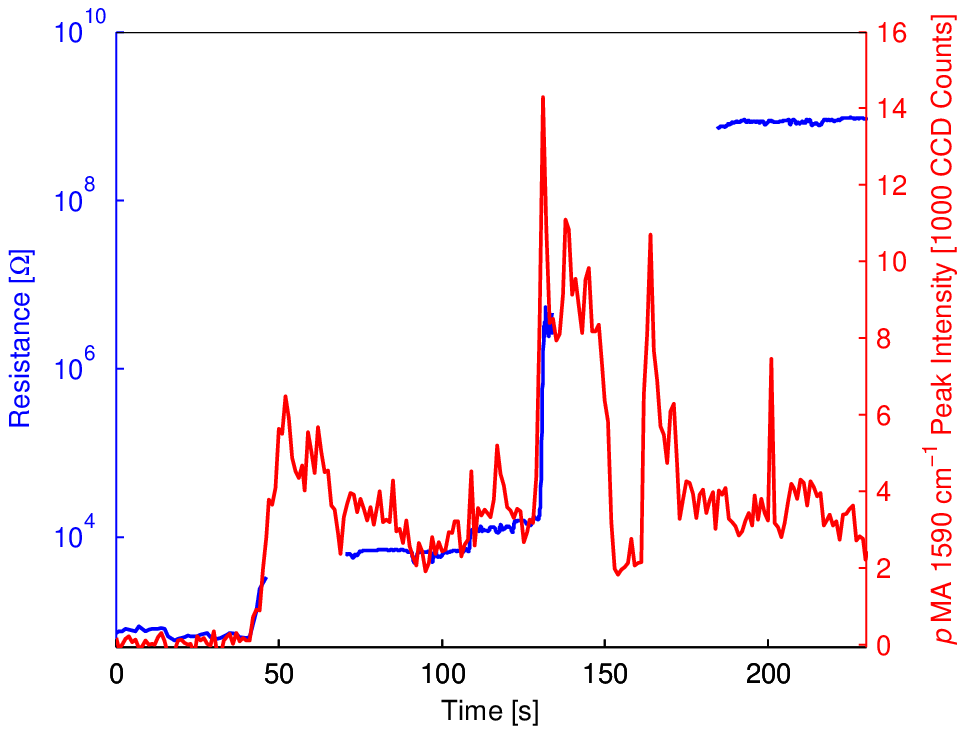}
\end{center}
\caption{
Blue curve (left scale): Resistance as a function of time for a different nanogap as it is migrated.  Breaks in the curve occur
where the gain of the current amplifier was adjusted to maintain signal.  Red curve (right scale):  CCD counts per second in
the 1590~cm$^{-1}$ peak (integrated from 1550-1650~cm$^{-1}$) of \textit{p}MA as a function of time for the same device
(synchronized with resistance plot).  The intensity of the peak increases linearly with the log of the resistance until the
resistance reaches around $10^6~\Omega$ where the Raman intensity then drops off significantly and no longer correlates with
changes in the resistance.  Stochastic intensity fluctuations (``blinking'') are observed beyond this point.
}
\end{figure}

Figures S3-S5 are additional examples of \textit{p}MA Raman spectra
that correlate in time with measured junction conductance.  Figure S6 shows
similar data from one FOPE device and another \textit{p}MA device.  Sometimes
the relationship between SERS intensity and $G(t)$ is simple, as in
Fig. S3A and S4A: increased Raman intensity correlates with increased
conductance.  However, in some devices {\it decreases} in SERS
intensity, particularly of $b_{2}$ modes, correlates with increased
conductance, as in Figs. S3B and S4B.  More complicated situations can
also arise, as in Fig. S5, where conductance changes appear to
correlate with both changes in Raman intensity and with the onset of
spectral diffusion of some modes.  These data contain much information
implicit about the orientation of the molecule, its bonding to the
metal atoms of the electrodes, and the local chemical environment.  A
realistic theoretical treatment that encompasses these necessary
ingredients is beyond the scope of this paper.  

\begin{figure}[ht]
\begin{center}
\includegraphics[clip, width=6in]{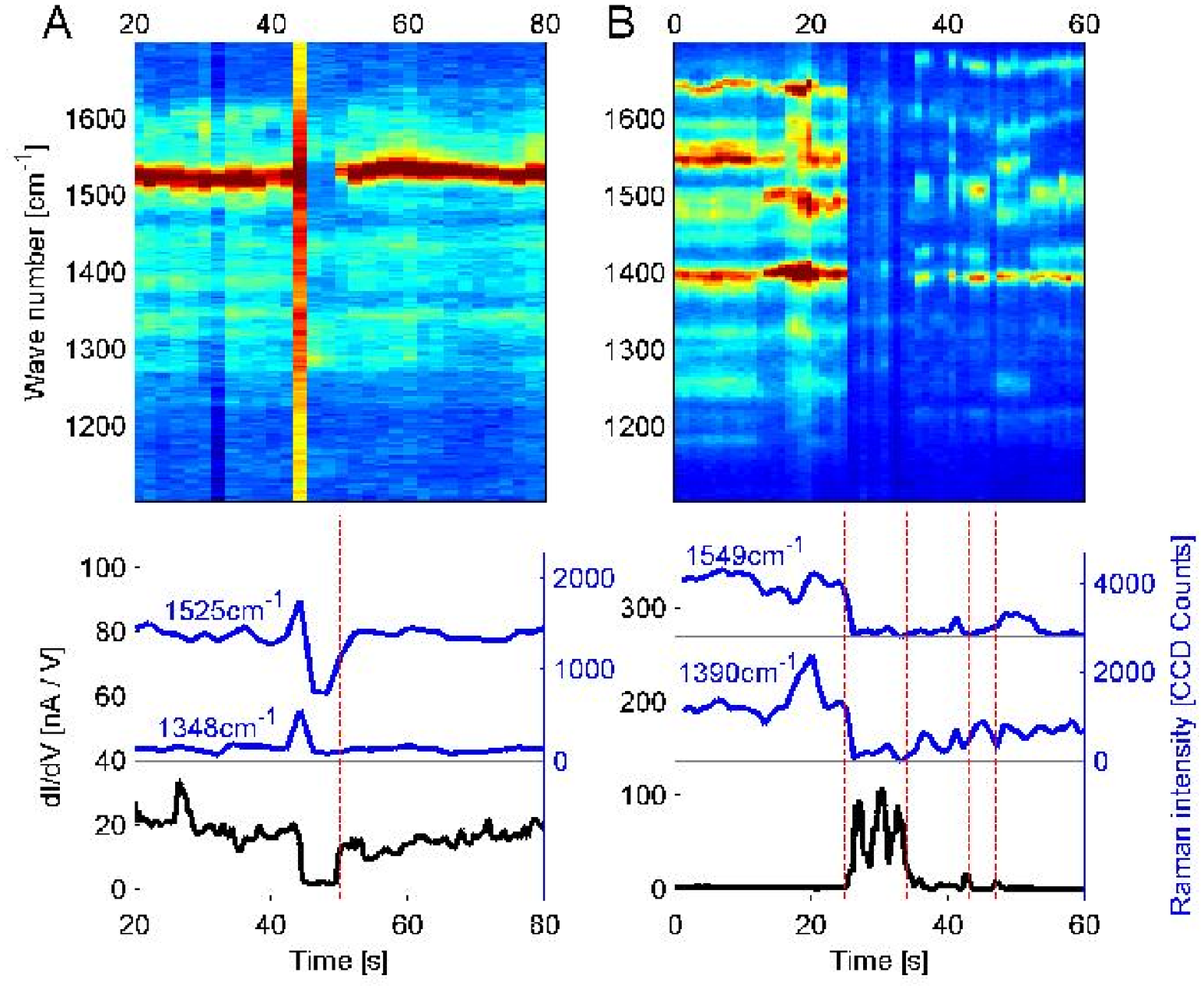}
\end{center}
\caption{
(A) Waterfall plot of Raman spectrum (2~s integrations) and positively correlated differential
conductance measurements of a \textit{p}MA sample (dark blue = 50 counts; dark red = 160 counts). Only the 1525~cm$^{-1}$ Raman mode exhibits a correlation with the conduction.  The other modes as illustrated by the 1348~cm$^{-1}$ mode do not show any changes expect at around 44~s where there is a temporary jump in the overall background. The 1525~cm$^{-1}$ mode has been shifted upward on the lower graph with the gray line indicating zero CCD counts. Vertical red lines indicate points of rapidly change Raman intensity and conduction. 
(B) Waterfall plot of Raman spectrum (1~s integrations) and negatively correlated conductance measurement for a different \textit{p}MA sample (dark blue = 0 counts; dark red = 250 counts). As illustrated by the modes at 1390~cm$^{-1}$ and 1549~cm$^{-1}$ conduction is only observed when the Raman intensity is almost undetectable.  The 1549~cm$^{-1}$ mode has been shifted upward on the lower graph for clarity.
}
\end{figure}

\begin{figure}[ht]
\begin{center}
\includegraphics[clip, width=6in]{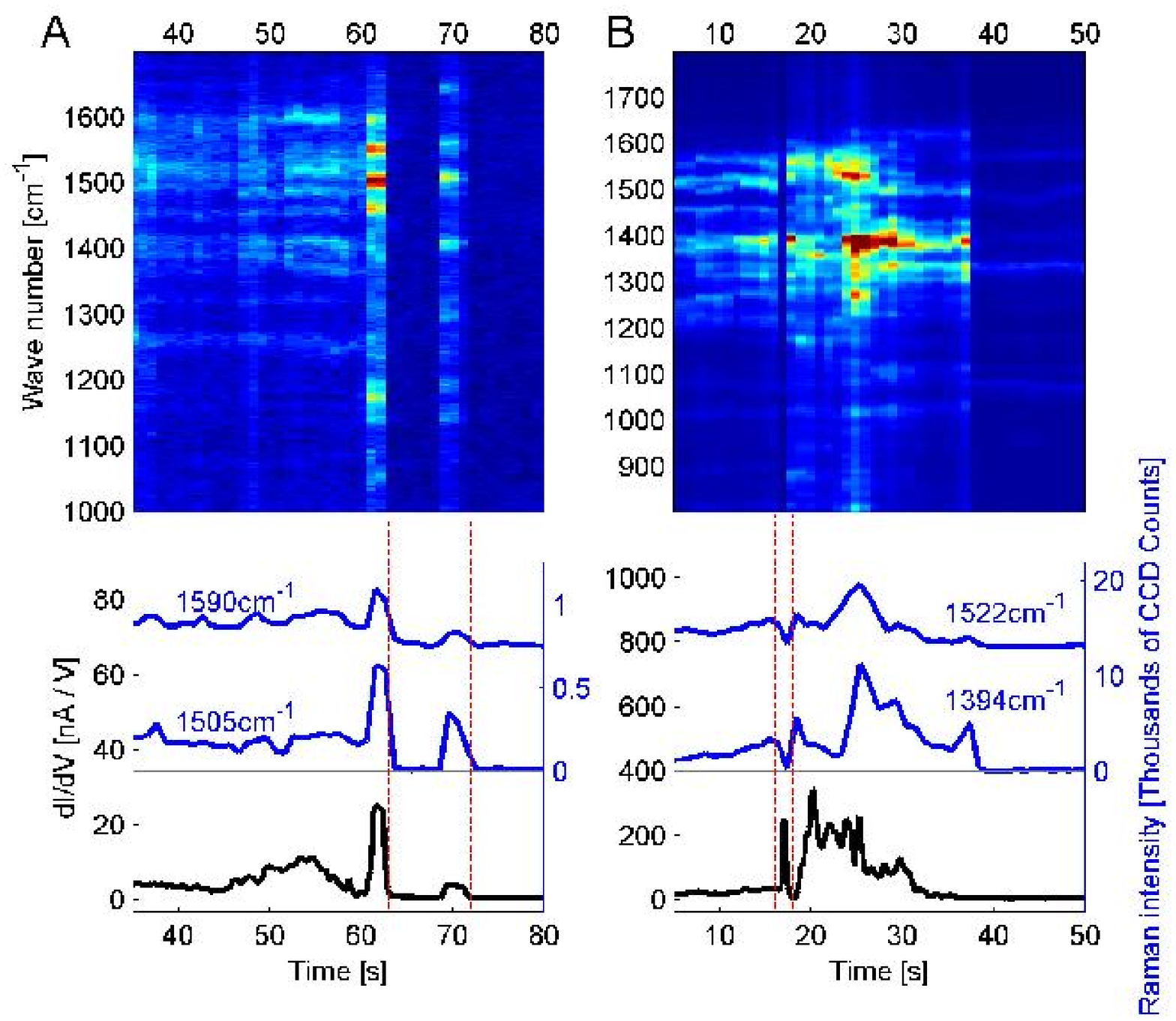}
\end{center}
\caption{
(A) Waterfall plot of Raman spectrum (1~s integrations) and positively correlated differential
conductance measurements for a \textit{p}MA sample (dark blue = 10 counts; dark red = 110 counts). For the most part the Raman intensity is very weak, however the intensity grows between 50 and 60~s culminating in a strong ``blink'' on and then off for the entire Raman spectrum.  This is followed be an additional blink at around 70 s.  Both the 1505~cm$^{-1}$ and 1590~cm$^{-1}$ modes show a  correlation with the conduction during this period of blinking. The 1590~cm$^{-1}$ mode has been shifted upward on the lower graph with the gray line indicating zero CCD counts.
(B) Waterfall plot of Raman spectrum (1~s integrations) and correlated conductance measurement for a different \textit{p}MA sample (dark blue = 20 counts; dark red = 1000 counts). This sample experiences a more complicated correlations as seen in the at 1394~cm$^{-1}$ and 1522~cm$^{-1}$ modes.  A spike in conductance is observed when the Raman spectrum completely disappears at around 18 s.  The conduction and Raman intensity then fluctuate in a negatively correlated way. Finally at 30~s the conduction drops off while the mode at 1394~cm$^{-1}$ continues to change. The 1522~cm$^{-1}$ mode has been shifted upward on the lower graph for clarity.
}
\end{figure}

\begin{figure}[ht]
\begin{center}
\includegraphics[clip, width=6in]{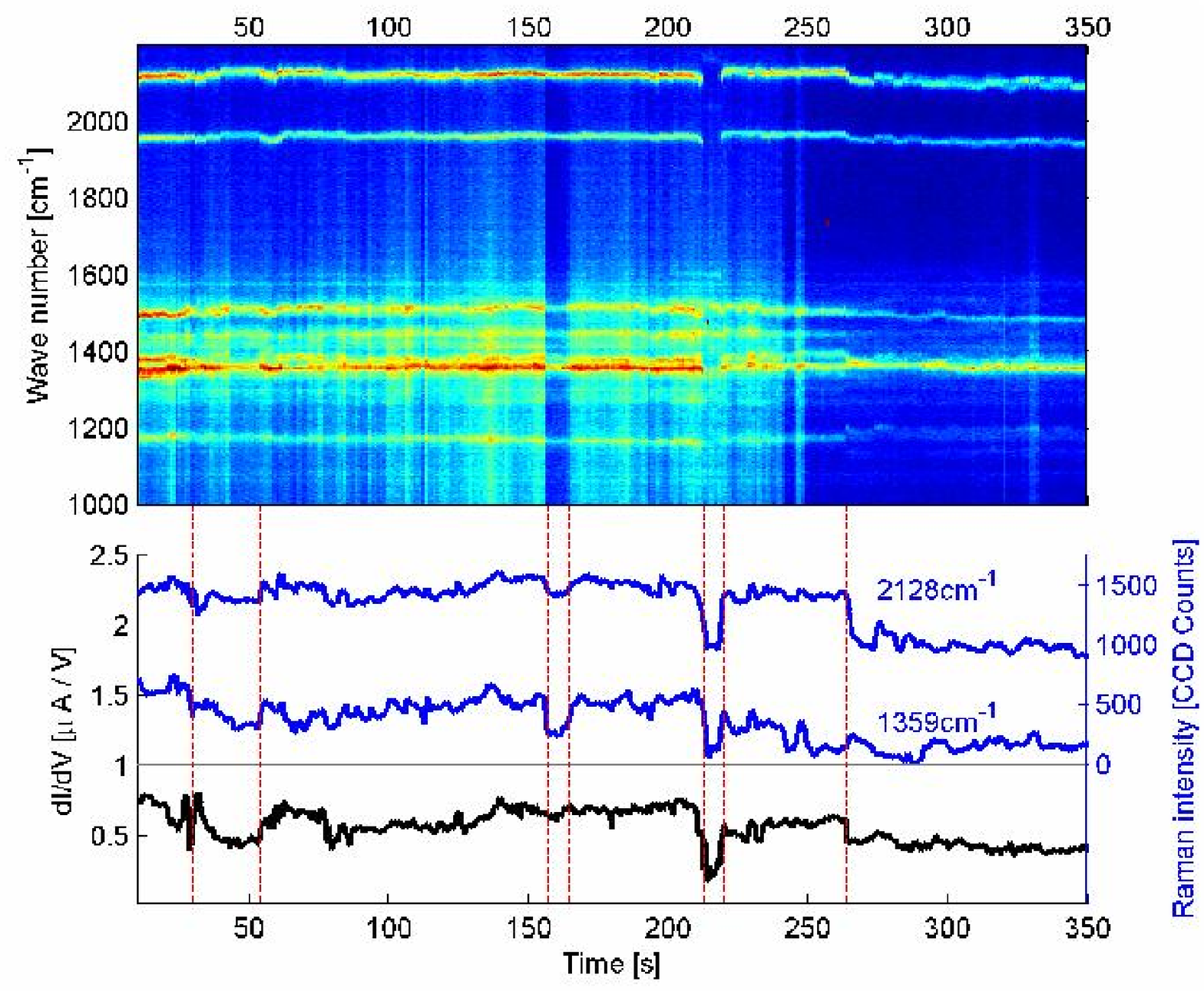}
\end{center}
\caption{
Waterfall plot of Raman spectrum (1~s integrations) and conduction measurements for another \textit{p}MA sample. This device shows that the conduction may be correlated with spectral wandering rather than intensity changes in the Raman spectrum as is evident in the 2128~cm$^{-1}$ mode.  We observe spectral wandering changes between 25 and 55~s that appear to correlate with alterations in the conduction.  At 160~s a drop in the overall background is observed that artificially lowers the intensity of all modes but has no effect on conduction.  At 213~s the mode at 2128~cm$^{-1}$ along with others vanishes and the conduction also drops significantly and returns along with the Raman modes at 220 s.  Finally a strong shift in wavenumber for the peak at 2128~cm$^{-1}$ is observed at 265~s corresponding with a drop in conduction at the same time.  The 2128~cm$^{-1}$ mode has been shifted upward on the lower graph for clarity.
}
\end{figure}

\begin{figure}[ht]
\begin{center}
\includegraphics[clip, width=6in]{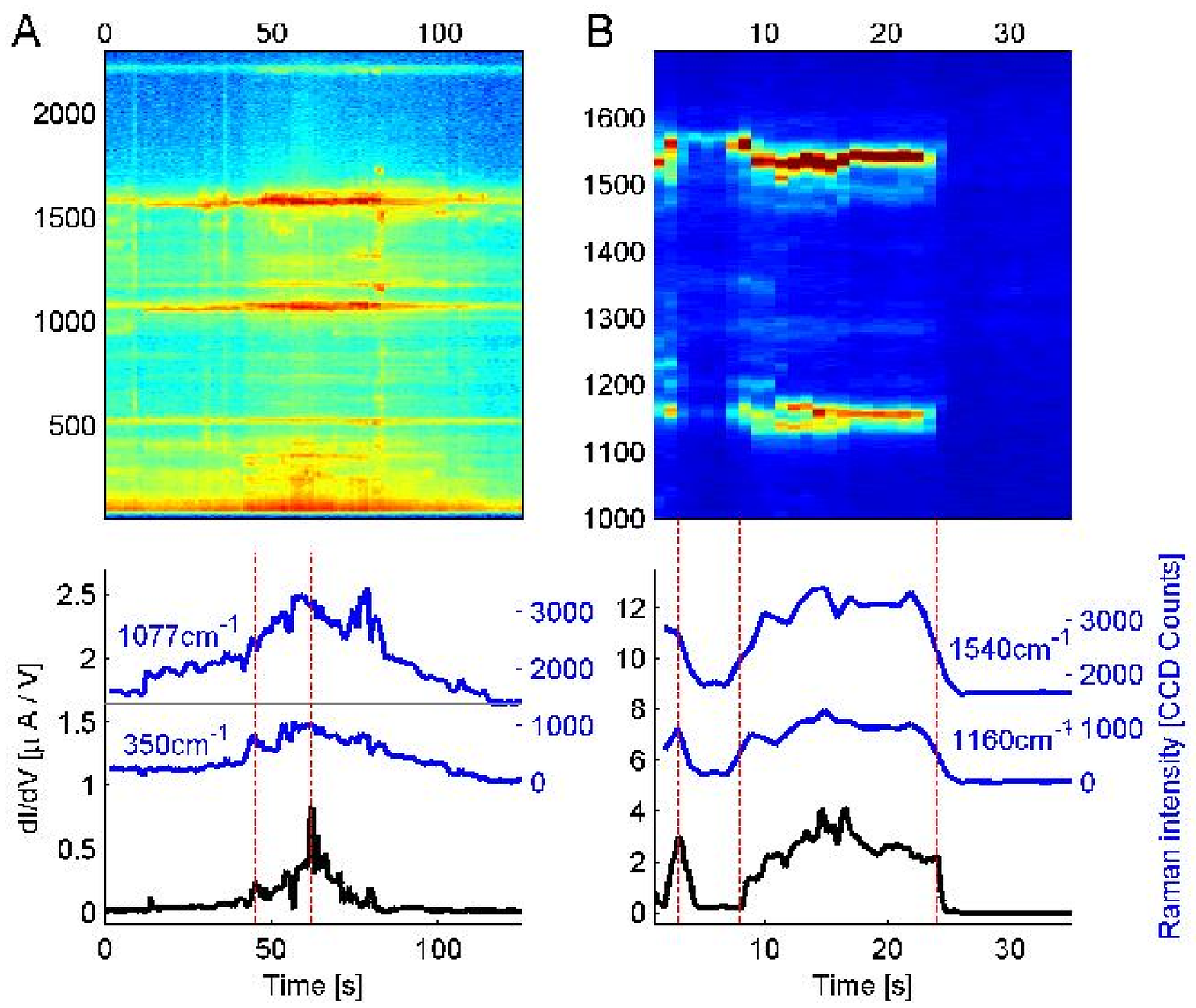}
\end{center}
\caption{(A) Waterfall plot of Raman spectrum (1 s integrations) and positively correlated differential conductance measurements for FOPE molecule.  The Raman intensity is plotted on a log scale for clarity(dark blue = 4 counts; dark red = 410 counts).  The Raman intensity for the 350~cm$^{-1}$ and 1077~cm$^{-1}$ is observed to steadily peaking at 62~s for both modes and again at 80~s for the 1077~cm$^{-1}$ mode.  The conductance also peaks at 62~s but also has some obvious differences in line shape from the Raman intensity.  However, the log of the conduction shows good correlation with the 350 cm$^{-1}$ peak.  The 1077~cm$^{-1}$ mode has been shifted upward on the lower graph with the gray line indicating zero CCD counts.  The slower response of the Raman spectrum compared to the conductance is due to the relatively slow integration time.  (B) Waterfall plot of Raman spectrum (1 s integrations) and positively correlated conductance measurement for a \textit{p}MA device (dark blue = 60 of fewer counts; dark red = 200 counts). Strong spectral wondering is observed with no correlation to changes in conductance. Both visible modes at 1160~cm$^{-1}$ and 1540~cm$^{-1}$ are positively correlated, no a1 symmetry modes are visible. The slower response of the Raman spectrum compared to the conductance is due to the relatively slow integration time. The 1540~cm$^{-1}$ mode has been shifted upward on the lower graph for clarity.
}
\end{figure}

\section{Continuum emission at low wavenumbers}

As mentioned in the text, in these electromigrated junctions there is
enhanced continuum Raman emission at low wavenumbers.  This emission
results from inelastic light scattering via the conduction electrons
in the metal, and is present in junctions without molecules.  Note
that in devices with molecules, in extremely strong SERS blinks (brief
periods of strong Raman emission), some increase in emission at low
wavenumbers does occur, though this is much more modest than the
percentage changes seen molecule-specific Raman modes.  Figures S7 and
S8 are reproductions of the Raman data from Figures 3 and 4 of the
main manuscript, but with the conductance as a function of time
plotted in comparison to the integrated continuum emission
(70~cm$^{-1}$ - 100~cm$^{-1}$).  During the major conductance changes
and Raman blinking events, the continuum emission changes much less
than either the SERS intensity or the conductance.  This strongly
suggests that the metal configuration at the junction is not changing
in time; rather, the changes in SERS intensity and conductance are
therefore due to changes in the molecular position, orientation, and
chemical environment.

\begin{figure}[ht]
\begin{center}
\includegraphics[clip, width=6in]{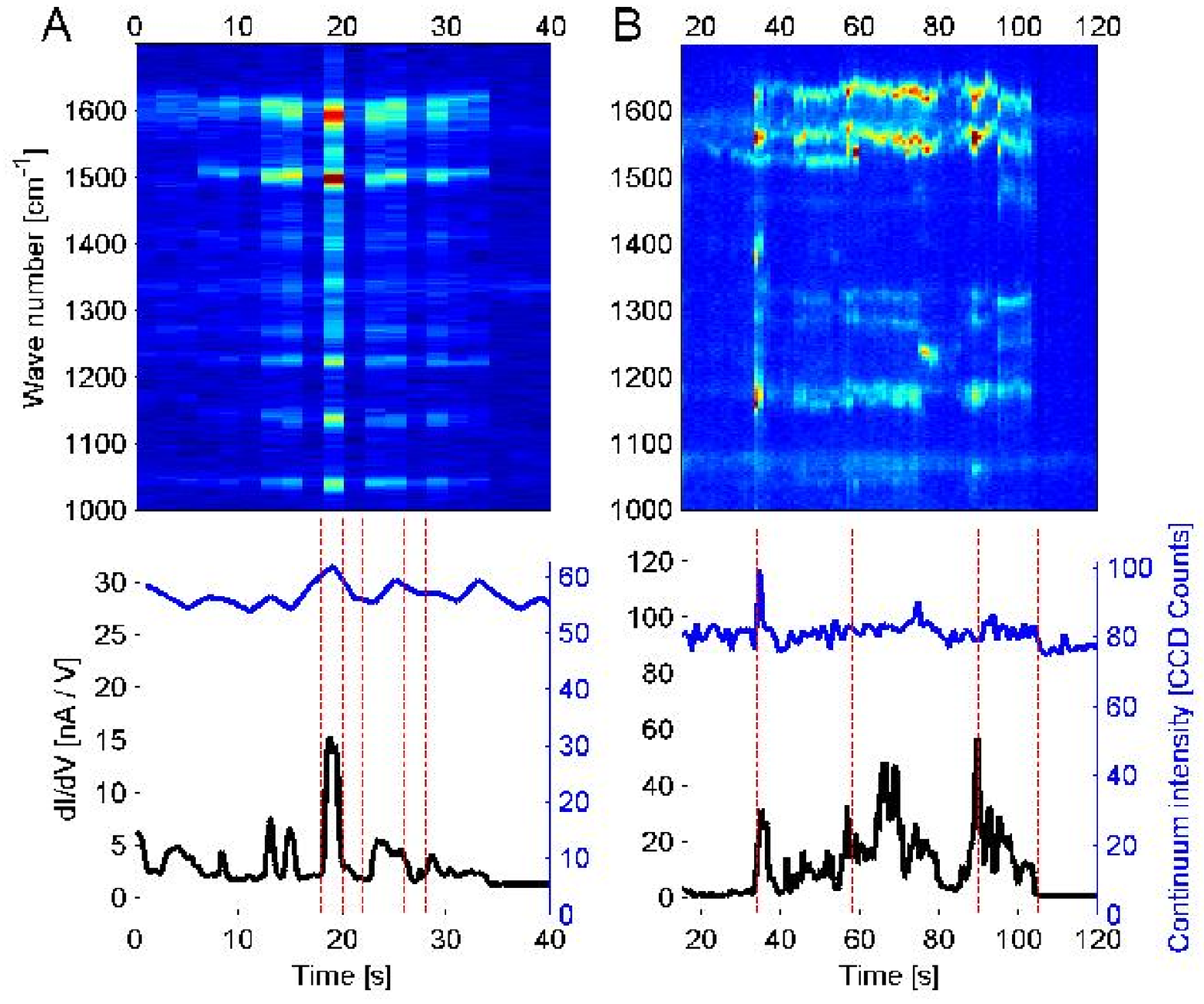}
\end{center}
\caption{
Reproduction of figure 3 from original paper.  Here the low wavenumber continuum intensity (integrated from 70-100~cm$^{-1}$)
is plotted as a function of time.  In both A and B the continuum is observed to be relatively constant with fluctuations much
smaller than those seen in the Raman modes.  
}
\end{figure}

\begin{figure}[ht]
\begin{center}
\includegraphics[clip, width=6in]{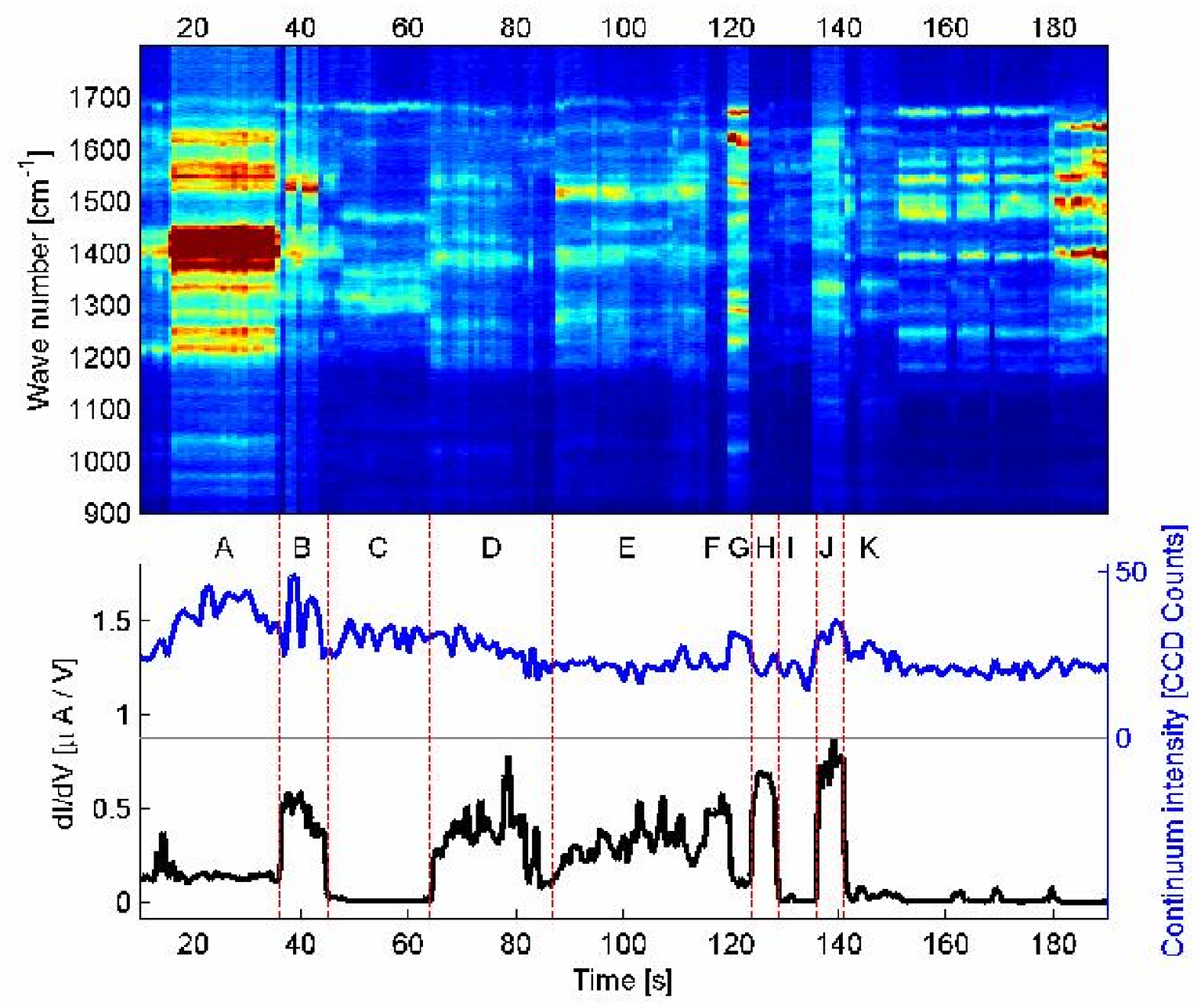}
\end{center}
\caption{
Reproduction of figure 4 from original paper.  Here the low wavenumber continuum intensity (integrated from 70-100~cm$^{-1}$)
is plotted as a function of time.  The continuum shows little correlation with the overall conduction especially when compared with the strong correlation exhibited by the active Raman modes.  The observed jumps in continuum typically correspond with times when the background of the entire spectrum is elevated relative to very high wave numbers ($>$~3000~cm$^{-1}$).  This could be the result of continuum generation from the molecule.  
}
\end{figure}

\clearpage

\section{FOPE molecule}
Figure S9 shows the Raman spectrum for a bulk crystal of the FOPE
molecule, first reported by Hamadani \textit{et
  al.}[S1]   The mode at approximately
2210~cm$^{-1}$ is the symmetric stretch of the carbon-carbon triple
bond.  The full name of FOPE is thioacetic acid
4-pentafluorophenylethynyl-phenyl ester, and its synthesis and
characterization are reported in the supplemental material to the
above paper.

\begin{figure}[ht]
\begin{center}
\includegraphics[clip, width=6in]{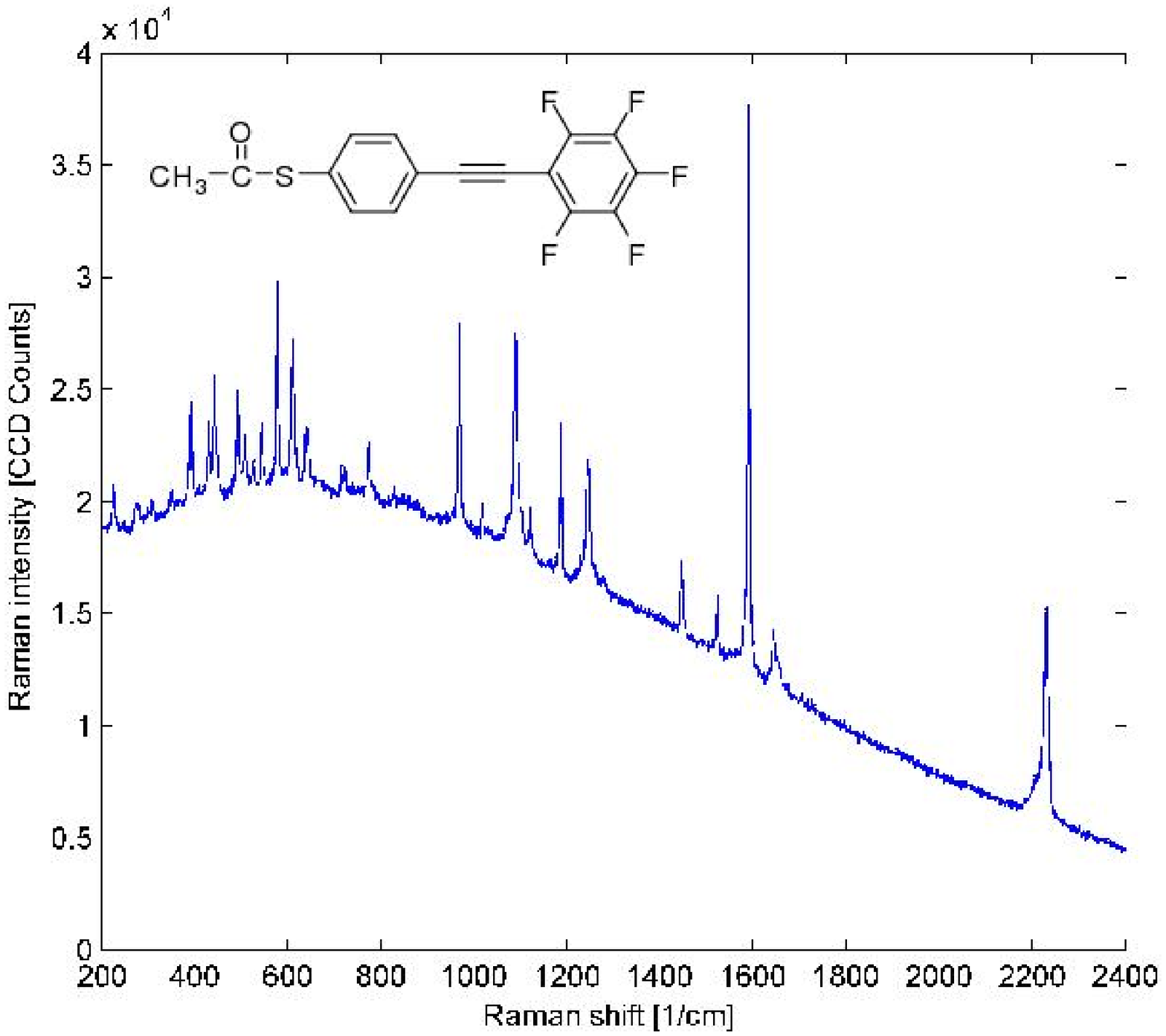}
\end{center}
\caption{Raman spectrum of bulk FOPE crystal taken in Renishaw Raman microscope at 785 nm.  Curvature and offset of spectrum is due to fluorescence.
}
\end{figure}

\clearpage

\section{FDTD calculations}
The optical properties of the bowtie structure were calculated using
the Finite-Difference Time-Domain method (FDTD) using a Drude
dielectric function with parameters fitted to the experimental data
for gold. This fit provides an accurate description of the optical
properties of gold for wavelengths larger than
500~nm.[S2] These calculations do not account
for reduced carrier mean free path due to surface scattering in the
metal film, but such an effect is unlikely to change the results
significantly.

The bowtie is modeled as a two finite triangular structures as
partially illustrated in Fig.~5, left inset, of the manuscript. Our
computational method requires the nanostructures to be modeled to be
of finite extent. The plasmon modes of a finite system are standing
modes with frequencies determined by the size of the sample and the
number of nodes of the surface charge distribution associated with the
plasmon. For an extended system such as the bowties manufactured in
this study, the plasmon resonances can be characterized as traveling
surface waves with a continuous distribution of wavevectors.

Previous calculations of a series of calculations of bowties with
increasing length reveals that the optical spectrum is characterized
by increasingly densely spaced plasmon resonances in the wavelength
regime 500-1000~nm and a low energy finite-size induced split-off
state involving plasmons localized on the outer surfaces of the
bowtie. For a large bowtie, we expect the plasmon resonances in the
500-1000~nm wavelength interval to form a continuous
band.[S3,S4]

To investigate the effects of a conductance shunting the nanoscale
gap, FDTD calculations were performed for a bowtie with two
semi-spherical protrusions in the junction as shown in the close-up in
the inset of Fig.~5 of the main text.  The electrodes are modeled as regular trapezoids of a height of 50~nm, and a 1~nm grid size was used.  The
conductance was modeled as a cubical volume 2~nm on a side located
between the 6~nm radius asperity and the facing electrode, where the
local field enhancement is maximized for modes relevant to the
wavelengths used in the experiment.  The conductivity of the material
was set to be frequency independent over the wavelength range of
interest (as expected for tunneling), and chosen such that the
conductance of that interelectrode link was the desired value.

Figure S10(top) reproduces the main portion of Fig.~5 from the main
text.  The three most prominent features in the calculated extinction
spectrum are labeled.  Peak ``a'' corresponds to the mode shown in
Fig.~5 of the main text, believed to be most relevant for the
experiments at hand.  Figure S10(bottom) shows the evolution of the
electric field enhancement factor (calculated for the mid-point of the
conducting volume standing in for the molecule) as a function of
interelectrode conductance.  The field enhancement and mode shapes due
to local features in the junction are essentially unaffected by the
interelectrode conductance until that conductance exceeds the order of
$G_{0}\equiv 2e^{2}/h = 7.74 \times 10^{-5}$~S.  For conductances
significantly larger than $G_{0}$, charges can flow between the two
electrodes and a new low energy plasmon resonance appears at
wavelengths that depend on the conductance of the shunt.  This is
completely consistent with the observations in Fig.~2 of the main
paper and Fig. S1 of this supporting material.  This demonstrates that
the conductance changes in these experiments are not able to modify
Raman emission from other molecules not in the tunneling region.

\begin{figure}[ht]
\begin{center}
\includegraphics[width=5in]{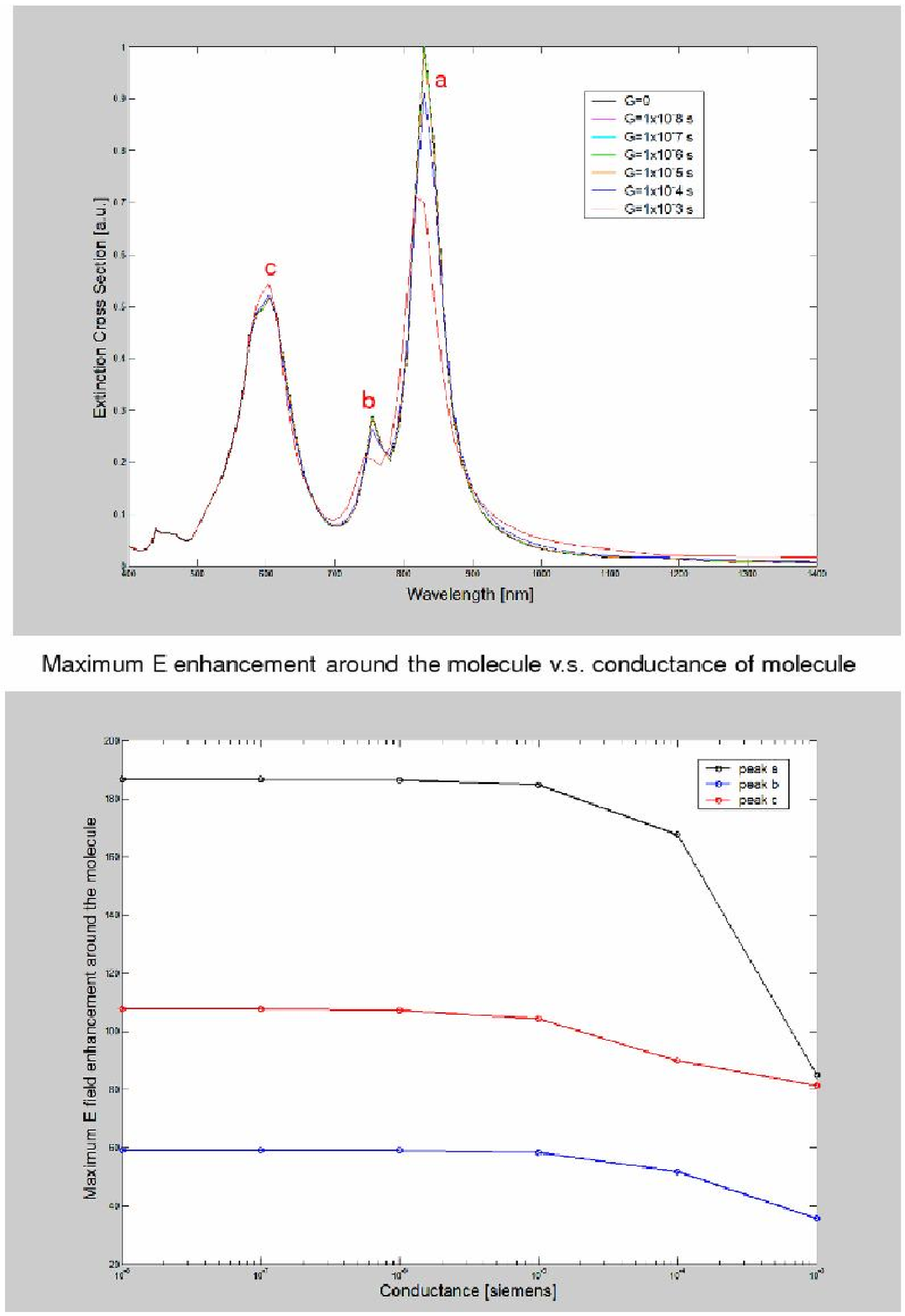}
\end{center}
\caption{Top:  Extinction spectra via FDTD calculations as explained
in the main text.  Bottom:  Field enhancement (relative to incident 
field) calculated at the midpoint of the interelectrode conductance
region for the three modes labeled in the upper graph, as a function
of interelectrode conductance.  Very little change in field
enhancement or distribution is seen until interelectrode 
conductance exceeds $G_{0}$.
}
\end{figure}

\clearpage

%%\begin{thebibliography}{10}

%%\bibitem{HamadanietAl06NL}
\noindent [S1] Hamadani, B. H.; Corley, D. A.; Ciszek, J. W.; Tour, J. M.; Natelson, D.  {\it Nano Lett.\/} {\bf 2006}, {\it 6}, 1303-1306.

%%\bibitem{OubreNordlander04JPCB}
\noindent [S2] Oubre, C.; Nordlander, P. {\it J. Phys. Chem. B\/} {\bf 108}, {\it 108},  17740-17747.

%%\bibitem{NordlanderLe06APB}
\noindent [S3] Nordlander, P.; Le, F. {\it Appl. Phys. B\/} {\bf 2006}, {\it 84}, 35-41.

%%\bibitem{WardetAl07NL}
\noindent [S4] Ward, D. R.; Grady, N. K.; Levin, C. S.; Halas, N. J.; Wu, Y.; Nordlander, P.; Natelson, D.  {\it Nano Lett.\/} {\bf 2007}, {\it 7}, 1396-1400.

\end{document}